\documentclass[a4paper,fleqn,usenatbib,useAMS]{mnras}

\usepackage{graphicx}   
\usepackage{amsmath}    
\usepackage{amssymb}    
\usepackage{multicol}        
\usepackage{bm}         
\usepackage{pdflscape}  

\usepackage[T1]{fontenc}
\usepackage{ae,aecompl}

\usepackage{newtxtext,newtxmath}

\title[Jet Properties of XTE~J1752-223]{Jet Properties of XTE J1752-223 During its $2009-2010$ Outburst}
\author[Debnath et al.]{Dipak Debnath$^1\thanks{E-mail: dipakcsp@gmail.com}$, Kaushik Chatterjee$^1\thanks{E-mail: mails.kc.physics@gmail.com}$, Debjit Chatterjee$^{2,1}$, Arghajit Jana$^{3,1}$, 
\newauthor{Sandip K. Chakrabarti$^{1}$}\\
\\
$^1$ Indian Centre For Space Physics, 43 Chalantika, Garia Station Road, Kolkata, 700084, India\\
$^2$ Indian Institute of Astrophysics, Koramangala, Bengaluru, Karnataka 560034    \\
$^3$ Physical Research Laboratory, Navrangpura, Ahmedabad 380009, India}

\date{Accepted 2021 April 21. Received 2021 April 21; in original form 2020 September 1}

\pubyear{2021}

\begin{document}
\maketitle

\begin{abstract}

Galactic short orbital period black hole candidate (BHC) XTE~J1752-223 was discovered on 2009 Oct 21 by the Rossi X-ray Timing Explorer (RXTE).
We study the spectral properties of this outburst using transonic flow solution based two component advective flow (TCAF) model.
TCAF model fitted spectrum gives an estimation of the physical flow parameters, such as the Keplerian disk rate, sub-Keplerian halo rate,
properties of the so-called {\it{Compton cloud}}, other than the mass of the source and normalization ($N$). $N$ is a standardized ratio
of emitted to observed photon flux in TCAF which does not include X-ray emission from jets. In the presence of jets, this ratio changes and
this deviation is used to obtain the estimation of X-ray contribution from the jets. Nature of the jet is found to be compact 
during low luminous hard state and discrete or blobby during high luminous intermediate states. We find a correlation between
the radio (5.5 GHz) and X-ray ($2.5-25$ keV) fluxes from different components. The radio ($F_R$) and jet X-ray ($F_{ouf}$) fluxes are found
to be correlated within the acceptable range of the standard correlation ($0.6$ to $0.7$). A similar correlation indices were reported by our group 
for three other short orbital period transient BHCs (Swift~J1753.5-0127, MAXI~J1836-194 \& XTE~J1118+480).

\end{abstract}

\begin{keywords}
X-rays: binaries -- stars individual: (XTE~J1752-223) -- stars:black holes -- accretion, accretion disks -- radiation:dynamics -- ISM: jets and outflows
\end{keywords}

\section{Introduction}

Black holes (BHs) are the end products of stars. Generally stellar massive black hole candidates (BHCs) reside in binary systems, and they occasionally 
go through outbursting phases of duration ranging from weeks to months. In between two outbursts they may stay in a long period of inactivity i.e., quiescence 
phase. During an outburst, electromagnetic radiation comes out from the accretion disk around the BH, which varies from radio to $\gamma$-ray. The radiation 
spectrum of a BH consists of two types of components: a soft multi-color disk black-body (DBB) and a hard power-law (PL) component. The origin of the soft component 
is the optically thick and geometrically thin Keplerian disk or Shakura-Sunyaev standard disk (Shakura \& Sunyaev 1973) and the hard component is believed 
to originate from a hot Compton cloud. Over the years, many models were put forward to explain the spectra of a BH. It is usual to fit a spectrum using a 
multi-color black body and a power-law component. A physical model in this context has been put forward by Chakrabarti and his collaborators which is based 
on viscous transonic flow solution which includes radiative transfer (see, Chakrabarti 1996a; 1995; Chakrabarti \& Titarchuk 1995, hereafter CT95; Chakrabarti 
1997). This so-called two component advective flow or TCAF solution consists of two components: a high viscous Keplerian component with high angular momentum, 
low radial velocity and a low viscous sub-Keplerian component with low angular momentum and higher radial velocity than the Keplerian component. TCAF model 
successfully describes the spectra of any BH, galactic or extra-galactic, where the Keplerian component produces the soft component and the sub-Keplerian 
component produces a hard component by processing some of the intercepted photons from the Keplerian disk. During an outburst, a BH goes through various spectral 
states (see, Debnath et al. 2015, 2017 and references therein). A black hole usually passes through four spectral states during a complete outburst: hard state 
(HS), hard intermediate state (HIMS), soft intermediate state (SIMS) and soft state (SS) (see, Nandi et al. 2012; Debnath et al. 2013). In an outburst, a 
classical or type-I BHC starts the outburst in the HS, then makes the transition towards the SS via HIMS and SIMS in the rising phase. After attaining the SS it 
again goes back to HS via SIMS and HIMS in the declining phase (Debnath et al. 2013, 2017 and references therein). Other type-II or harder types of outbursts 
do not show SS or even a SIMS. In the TCAF paradigm, this is due to the dominance of the sub-Keplerian halo rates both via accretion and winds from the companion. 
Low frequency quasi-periodic oscillations (LFQPOs) are also common phenomena in BH outbursts (Remillard \& McClintock, 2006). It is observed that type-C 
QPOs show a monotonic increase and decrease in frequency during HS and HIMS of rising and declining phases of an outburst respectively (Debnath et al. 2008, 2013; 
Nandi et al. 2012). In SIMS, generally, type-A, or type-B LFQPOs are observed sporadically (Nandi et al. 2012). SS does not show any signature of LFQPO. 

Jets/outflows are very important phenomena in BHs. Although jets are common in AGNs, some of the stellar massive BHs also show this high energetic phenomenon. 
In astrophysical jets, the ionized matter is emitted as a beam along the axis of rotation with which mass, energy, momentum are channeled from stellar, Galactic 
and extra-galactic BHs along the axis. Jets are geometrically narrow and conical. The most powerful jets are associated with AGNs. Though there is diversity 
in luminosity and other properties, the structure (morphology) of jets from SBHs and AGNs are similar. While intensive radio observations of BHCs originally 
discovered compact jets, later it became clear that jets emit radiation in a broad range from radio to $\gamma$-rays due to synchrotron radiation. Compact 
radio jets are observed in Cyg~X-1 (Stirling et al. 2001), GRS~1915+105 (Dhawan et al. 2000), while high energy $\gamma$-ray jets have been observed in Cyg~X-1 
(Laurent et al. 2011; Jourdain et al. 2012), V 404 Cyg (Loh et al. 2016). There is also a spectral break in near-infrared frequency (Corbel \& Fender 2002) 
at which transitions from optically thick to optically thin synchrotron radiation occurs. This spectral break has been observed in many BH X-ray binaries 
(BHXRBs) e.g., GX~339-4 (Corbel \& Fender 2002).

Though there is still debate on how jets are produced, over the years many models have been proposed to explain the origin, acceleration and collimation of 
the jets. The de-Laval nozzle model (Blandford \& Rees 1974), electrodynamically acceleration model (Znajek 1978), centrifugally driven outflow (Blandford 
\& Payne 1982), Blandford-Znajek mechanism (Blandford \& Znajek 1977) are some of the models which tried to explain the production and collimation of jets. 
In the TCAF model, the CENBOL or the CENtrifugal pressure supported BOundary Layer (CT95; Chakrabarti 1997) acts as the base of the jet. Here the radiation 
pressure is responsible for launching the jet (Chakrabarti 1999a; Das \& Chakrabarti 1999). The pre-jet flow is hot and subsonic close to a BH and can emit 
X-rays. After crossing the sonic surface (located at, say, $r=r_c$), it becomes supersonic. Chakrabarti (1998) first established the relation between outflow 
and inflow rates from the first principle. According to the TCAF model, the outflow remains isothermal and subsonic up to the sonic surface ($\sim 2.5~X_s$, 
where `$X_s$' is the shock location, i.e., the size of the CENBOL) close to the BH by the deposition of momentum by hard photons as it expands and cools down. 
In case of high accretion rates, CENBOL is cooled down rapidly, thereby quenching the outflow (Chakrabarti 1998). Jets are mainly of two types: compact or 
continuous jet and discrete or blobby jet (Chakrabarti \& Nandi 2000 and references therein). In the case of HS, when the CENBOL is very hot and big in size, 
compact jets are produced. In the intermediate (HIMS and SIMS) states, when CENBOL is very close to the BH, discrete or blobby jets may be produced. In the 
soft state, when the CENBOL is quenched, no jet is observed.

Can the jet emission contribute to the observed X-ray flux? Hannikainen et al. (1998) first pointed out for the BHC GX~339-4 that the radio and X-ray emission 
are strongly correlated in the low hard state (LHS). The correlation study was done in detail by Corbel et al. (2003, 2013) and Gallo et al. (2003). A correlation 
was established as $F_R \propto F_X^b$ where $F_R$, $F_X$ are the radio and X-ray fluxes respectively and $b$ is the correlation index. They found $b \sim 
0.6-0.7$ for many BHCs. This is called the `standard' correlation between radio and X-ray radiations in active jets. There are some BHCs (for e.g., Swift~J1753.5-0127, 
XTE~J1650-500, IGR~J17497-2821, MAXI~J1836-194, etc.), which follow a steeper correlation between radio and X-rays with a correlation index $b > 1.0$ 
(Jonker et al. 2004, Corbel et al. 2013, Jana et. al. 2017, 2020). These are called `outlier' sources. 

XTE~J1752-223 was discovered by the Rossi X-ray Timing Explorer (RXTE) on 2009 October 21. This source is situated in the galactic bulge, at R.A.=$268.05 \pm 
0.08$, Dec.=$-22.31 \pm 0.02$ (J2000 coordinate) with a distance of $d = 3.5 \pm 0.4$ kpc (Shaposhnikov et al. 2010). According to Ratti et al. (2012), it is a 
short orbital period ($\sim$ 6.8 hrs) transient BHC having a $M$ type donor companion star. It has a spin parameter of $a = 0.52 \pm 0.11$ (Reis et al. 2011) 
and an inclination angle $i < 49^{\circ}$ (Miller-Jones et al. 2011). This outburst was active for almost eight months. However, there was absence of RXTE PCA 
data from 2009 November 20 to 2010 January 19 due to the Sun constraint. According to Shaposhnikov et al. (2010), this source has gone through all the canonical 
spectral states of a BHC. In Chatterjee et al. (2020; hereafter Paper I), detailed spectral analysis of the source was carried out using both DBB+PL and TCAF 
models. They found that the source had gone through all four canonical spectral states in the following way: HS (rising) $\rightarrow$ HIMS (rising) $\rightarrow$ 
SIMS or SS $\rightarrow$ HIMS (declining) $\rightarrow$ HS (declining). 

During this entire outburst, radio jet was highly active as reported by many authors (see for examples, Brocksopp et al. 2009, 2013; Russell et al. 2012). Using 
Australia Telescope Compact Array (ATCA) observations, Brocksopp et al. (2009) reported a radio counterpart of 2 mJy in both 5.5 and 9 GHz bands. Using multi
wavelength observation, Russell et al. (2012) reported a late jet re-brightening in the decaying hard state. In general, during SS, there is no production of 
jets/outflows. In this outburst, the source was in SS in the middle phase of the outburst (Shaposhnikov et al. 2010), where Brocksopp et al. (2013) reported the 
existence of optically thin radio flares. According to TCAF, the size of the CENBOL gets quenched when the source goes to the SS. Since the jets are produced 
from the CENBOL, we do not observe any jets in the SS. However, if the accretion disk is magnetically dominated, blobby jets can be observed in the SS also due 
to the magnetic rubber band effect (Nandi et al. 2001). For radiation pressure dominated disk one could also see discrete ejection events away from the source. 
In Paper I, a detailed study on the spectral and the temporal properties of the source are done during this outburst to infer accretion flow dynamics of the 
source using RXTE, Swift and MAXI data. Since the source was highly active in jets, in this paper, we have estimated X-ray contribution from the jets or outflows 
based on the spectral analysis with the TCAF model. Here, we first follow the method presented in Jana, Chakrabarti \& Debnath (2017) and then a new method, 
which is introduced here after adding an additional power-law model (for jets) with the constant normalized TCAF model (for disk) to fit BH spectra. 
To understand the nature of the emitted jets, we have also studied the correlation between radio and X-ray from jets. 

The paper is organized in the following manner. In \S2, we briefly discuss about the disk-jet correlation method using TCAF method. In \S3, we discuss about
two methods of jet X-ray flux estimation. In \S4, the Observation and data analysis method is presented. In \S5, we present the results and in \S6, we give a 
brief discussion and concluding remarks. 

\section{Disk-Jet Connection, Spectral States and TCAF Solution}

In the TCAF model, one requires six parameters to fit the data out of which four are related to the flow properties, one is related to the black hole mass and 
the final one is related to the instrument parameters observing the black hole. These are: Keplerian disk rate (${\dot m}_d$), sub-Keplerian halo rate (${\dot m}_h$), 
shock location ($X_s$), compression ratio ($\rho$), mass of the BH ($M_{BH}$) and model normalization ($N$). The normalization ($N$) is a function of intrinsic source 
parameters such as the mass of the BH ($M_{BH}$), distance ($D$) and disk inclination angle ($i$), apart from the instrument area. So for a given object, TCAF does 
not have a provision to change $N$ across all the spectral states. On the other hand, if one requires significantly varying $N$ while fitting the data set of an 
outburst, that may indicate the presence of jet. This is because in TCAF the X-ray flux from the base of the jet was not included deliberately since there is no 
unique jet configuration for a given accretion flow configuration. When the jet is present, one requires higher $N$ to achieve the best fit as it compensates 
for extra X-ray emission from the base of the jets. 

Chakrabarti (1998) showed from purely hydrodynamic consideration that the jets are thermally driven from CENBOL, the outflow rate (${\dot M}_{out}$) is related 
to the inflow rate (${\dot M}_{in}$) as a function of the compression ratio ($R$). It follows the following relation: 
$$\frac{{\dot M}_{out}}{{\dot M}_{in}}={{\dot R}_m}={\frac{{\theta}_{out}}{{\theta}_{in}}}{\frac{R}{4}}\left[\frac{R^2}{R-1}\right]^{3/2}exp\left({\frac{3}{2}}-{\frac{R^2}{R-1}}\right) \eqno (1)$$
where ${\theta}_{out}$ and ${\theta}_{in}$ are the solid angles subtended by the outflowing and inflowing cones respectively. The compression ratio ($R=\rho_+
/\rho_-$, where $\rho_+$ and $\rho_-$ are post and pre-shock matter densities) varies between $\sim 1-4$. It becomes $\sim 1$ in SS and $\sim 4$ in HS and it 
stays in between in HIMS or SIMS.  The compression ratio $R$, in turn, will partly depend on other flow parameters, such as ${\dot m}_d$ and ${\dot m}_h$. 
According to this model, the jet moves subsonically up to the sonic surface, which is approximately 2.5 times the size of the CENBOL ($X_S$) and then moves 
away supersonically (Chakrabarti 1999a,b). The jet X-ray flux ($F_{ouf}$) is the net contribution by two processes: the upscattering of the seed photons 
from the Keplerian disk and downscattering of the CENBOL photons, both in the subsonic region of the jet where the optical depth is highest. $F_{ouf}$ does 
not take into consideration of the X-rays emitted by the interaction of the jet with the ambient medium. In the HS, the CENBOL size is big in the presence of 
a strong shock and is hotter than the incoming flow. `$R$' is high as well. This makes ${\dot R}_m$ small and compact jets come out from the CENBOL. In 
the HIMS, the CENBOL moves closer to the BH with intermediate shock strength. The CENBOL still remains hot (less hot compared to the HS) and `$R$' 
decreases. The outflow increases a bit as compared to the HS. In the SIMS, when the shock moves much closer to the BH and the supply of the Keplerian disk 
matter takes over the supply of the sub-Keplerian matter, the CENBOL cools much faster compared to the HS or the HIMS. In this state, the outflow becomes 
maximum with intermediate `$R$' and blobby jets may be observed. In the SS, when the CENBOL is totally cooled down by the inverse-Comptonization of seed 
photons and there is no shock, no jets/outflows can form, unless the inflow is super-Eddington and outflows are radiation pressure-driven as opposed to 
thermal pressure driven. The variation of ${\dot R}_m$ with $R$ is given in Fig. 3a of Chakrabarti (1999a). It shows that the outflow rate is low when any 
outburst starts in the HS, then it increases monotonically and reaches a maximum in the SIMS via HIMS and then it goes to zero in the SS. If the disk is 
magnetically dominated, the occurrence of optically thin radio flares may take place.

\section{Estimation of Jet X-ray Flux}

Although dominating radiation from jet is radio, it also emits radiation in a wide range of electromagnetic band. We see emission of the high energy 
X-ray radiation from the base of the jet. As jet moves outward, due to adiabatic expansion, matter density, temperature decreases to produce other low energetic 
radiation in the bands from UV, Optical, IR, to radio. In a jet dominated phase of a BH, the observed X-ray contains a contribution from two components: one from the 
accretion disk and the other from the jets. Jana et al. (2017; hereafter JCD17) tried to separate these two components of X-ray fluxes from total observed X-ray 
fluxes using TCAF model fitted constant normalization method. Other than this method, here we have used another method to separate disk and jet fluxes using 
spectral fit with the combined TCAF and PL models. Details of these two methods are discussed in the following sub-Sections.

\subsection{Constant Normalization Method using TCAF Model}

Recently Jana et al. (2017), using the fact that TCAF normalization ($N$) can vary with jet X-ray activity, separated total observed X-ray flux into 
two of its constituents (disk and jet) based on the spectral analysis using TCAF model. Unlike other models, TCAF normalization being a function of intrinsic 
parameters (mass, distance and inclination angle) does not vary on observations of a particular source (if observed with the same satellite instrument). 
A deviation of the constancy of the model parameter $N$ may be seen if there is any jet activity or any other dominating physical processes whose effects 
have not considered in the current version of the TCAF model fits file or there is a precession in disk, which actually changes effective area of the accretion 
disk. Since jet also emits X-ray, its emission adds up to the observed X-ray from the accretion disk. So, in the jet dominated phase of an outburst, we see higher 
$N$ values are required to fit spectra. This is because, extra N values (over the constant value) tries to compensate extra flux rise due to the excess X-ray 
contribution from jets. To confirm the excess X-ray is emitted from jets, JCD17  noticed that $N$ values followed similar trend as of the observed radio flux during 
2005 outburst of Swift~J1753.5-0127 and on the lowest N observed day, radio flux was also found to be at its lower range. So, they assumed on the lowest $N$ 
observed day, jet contribution in X-rays in minimum or negligible. They estimated disk or inflow flux ($F_{inf}$) using `flux err $E_{min}$ $E_{max}$' command 
after just putting/freezing model $N$ value at the lowest observed $N$ value in all best fitted observations. Actually they had not refitted observations with 
the lowest $N$ value, as it will change the model parameters and fit will be unsatisfactory. Actually, their goal was just to obtain flux contributions from the 
disk or inflow using lowest $N$ value, not to see the variation of model parameters with the frozen $N$ condition.

To separate disk and jet X-ray fluxes from total observed X-ray during the current outburst of XTE~J1752-223, we used similar method, as here also we found 
higher $N$ are required to fit spectra, when source was more active in radio (see, Fig. 1d, 1e). Similar to JCD17, here we also assumed that on the lowest 
$N$ observed day (on MJD=55371.9, i.e., 2010 Jun 24), source was inactive in jet. So, on that day, total observed X-rays are coming only from the accretion 
disk or inflowing matter. To estimate the flux contribution from accretion disk or inflowing matter ($F_{inf}$) in the PCA spectral energy range $2.5-25.0$~keV 
band, we freeze the normalization to the lowest $N$ value (obtained from the TCAF model fits with all parameters including $N$ as free condition) and run the task 
`\textit{flux err 2.5 25.0}'.

A model fit when all parameters including $N$ are kept free, gives us the total X-ray flux ($F_X$), and it includes X-ray contribution from both 
inflow or disk ($F_{inf}$) and outflow or jet ($F_{ouf}$), i.e., 
                                                          $$F_X = F_{inf} + F_{ouf},$$
or,                                                       $$F_{ouf} = F_X - F_{inf}.\eqno (2)$$
We thus obtain the X-ray flux contribution only from jets/outflows just by subtracting $F_{inf}$ from $F_X$, obtained by free $N$ and lowest frozen $N$ conditions 
respectively.

\subsection{Method of Constant Normalization TCAF (for $F_{inf}$) with an additional Power-law Model (for $F_{ouf}$)}

In active jet observations, we see excess contribution of X-ray flux on the top of the accretion disk X-ray. It is well established that the continuum 
of the disk X-ray contains two major components originated from thermal (multicolour black body or DBB shape) and nonthermal (PL shape). The nonthermal PL 
part mainly contains upscattered or inverse-Comptonized X-ray from the `hot' Compton cloud or CENBOL. Same Comptonization or Synchotron processes might be the 
primary processes for the generation of the jet X-ray at the base of the jet (up to sonic surface $\sim 2.5~X_s$). So, we may assume that the shape of the jet 
X-ray as a PL type having different slope index, compared to the disk component of the PL. From the comparative variation of the $N$ and $F_R$, we could assume 
that on the lowest $N$ observed day (if $F_R$ also stays at its lowest range) jet contribution in the observed X-ray is minimum or negligible. Based on this 
assumption, JCD17 calculated disk component of X-ray flux with the method as described above. In that method, $F_{inf}$ was obtained just by putting/freezing 
model $N$ value in all best fitted observations at the value of the lowest observed $N$ value of the entire period of the analysis. So in their method, refitting 
of the spectra was not done as it will change the fit parameters and statistics. Here, we have refitted all the spectra after adding a PL model with the TCAF 
model, considering constant normalized (at lowest $N$) TCAF model will take care of the disk X-ray contribution and PL model will take care of the jet component 
of the X-ray. Interestingly we found that TCAF model parameters was not changed significantly as obtained from the only TCAF model fit, where all model parameters 
(including $N$) were kept as free (see, Table 2 of Paper I and Table 2 of the present paper). After obtaining the best fit using TCAF+PL models, `flux err 2.5 
25.0' command was used to estimate $F_X$ in the $2.5-25.0$~keV PCA band, and same command was used to estimate $F_{inf}$ and $F_{ouf}$ fluxes using convolution 
model $`cflux'$ on TCAF and PL models respectively.

\section{Observation and Data Analysis}

We analysed archival data of 40 RXTE PCA observations \footnote{\url{https://heasarc.gsfc.nasa.gov/docs/xte/SOF/score.html}} 
from 2009 October 30 (MJD = 55134.11) to 2010 June 24 (MJD = 55371.95). Using the PCU2 data, we studied the source in $2.5-25$ keV energy range 
for all these observations. In Paper I, we showed the variation of accretion flow properties and estimated the mass of this source when all model 
parameters were kept as free. The mass of the source was obtained in the range $8.1-11.9~M_{\odot}$ with a probable mass of $10 \pm 1.9~M_{\odot}$. 
Based on the variation of the spectral fitted parameters and temporal properties, Chatterjee et al. (2020) found four major spectral states, 
namely, HS (rising) $\rightarrow$ HIMS (rising) $\rightarrow$ SIMS or SS $\rightarrow$ HIMS (declining) $\rightarrow$ HS (declining). 
An exact transition date between SIMS and SS could not be found and the central phase was termed as `SIMS or SS'. A broad range ($0.51-3.10$) 
of normalization ($N$) parameter was required to achieve the best fit suggesting the presence of strong X-rays from the jet. This is verified 
by comparing the variation of $N$ with $F_R$. In Fig. 1(c) and 1(d) we see a similar variation of these quantities. The situation is similar 
to the 2005 outburst of Swift~J1753.5-0127 (JCD17), 2000 outburst of XTE~J1118+480 (Chatterjee et al. 2019), 2011 outburst of 
MAXI~J1836-194 (Jana et al. 2020). 
Using both our radio vs X-ray correlation methods (mentioned in \S3), we have established correlations between radio ($F_R$) and different 
component of X-ray fluxes ($F_X$, $F_{inf}$, $F_{ouf}$). A tight correlation between $F_R$ and $F_{ouf}$ suggests the nature of the jet as compact 
and a loose correlation suggests that the nature of the jet to be discrete or blobby. For radio data, we use $5.5$ and $9$~GHz of ATCA data from 
Brocksopp et al. (2013) paper. Note here, $F_X$, $F_{inf}$ and $F_{ouf}$ are calculated in units of $10^{-9} erg~ cm^{-2}~ s^{-1}$ and $F_R$ is 
presented in $mJy$ unit.

\section{Results}

Results based on 40 PCA observations with the spectral fits using the TCAF model in the $2.5-25$ keV energy range are presented here. Based on the variation 
of the model normalization, we estimated X-ray flux contribution from jets/outflows. The fluxes are estimated using two methods as mentioned in \S 3.
To understand nature of the emitting jet, correlation between radio and jet X-ray fluxes are also studied. 

In Figs. 1 and 2, we show the variation of estimated X-ray fluxes ($F_X, F_{inf}$ and $F_{ouf}$) with TCAF model normalization ($N$) and the 
observed radio flux ($F_R$ in $5.5$~GHz ATCA data) from spectral analysis using only TCAF model, and TCAF plus PL models respectively.
Figures 3(a-d) and 4(a-d) show correlation plots between the radio and X-ray fluxes. 

\subsection{Evolution of the Jet Flux}

Jets are mainly observed in radio waves, although they also emit high energy radiation. X-rays are produced from the base of the jets mainly via Comptonization 
or synchrotron processes. Magnetic field plays an important role in the launching of the collimated jets or in the form of discrete blobs. As a jet moves away, 
due to adiabatic expansion, temperature drops and we observe low energy radiation i.e., UV, optical, IR and radio from the jets. During an outburst, the evolution 
of the jets occurs as accretion rate changes. In Figs. 1 and 2, we show the evolution of the observed radio flux and different components of X-ray fluxes as obtained 
from our analysis during the 2009-10 outburst of XTE~J1752-223. 

\subsubsection{Radio}

In Fig. 1(e) and 2(e), we show the variation of the radio flux of XTE~J1752-223 during the $2009-10$ outburst with 5.5 GHz ATCA data. These data have been 
adopted from the radio light curve of Brocksopp et al. (2013). The radio flux was in the lower range in the rising HS and reached its maximum on 2010 January 21 
(MJD = 55217.9), when the outburst was found to be in HIMS. This is the first radio peak out of the total seven peaks that the source showed during entire phase of the 
outburst (see, Figs. 1 \& 2). According to Brocksopp et al. (2013), the jet was compact during the harder states (both in the rising and declining phases) and as the 
source goes to the softer states, they reported the jet as discrete or blobby. The multiple radio peaks are also observed during the softer states (see Fig. 1e). 
The intensity of the radio peaks is also observed to decrease as the outburst progresses. This observation of radio flares is uncommon in the soft states. 
Normally, SS is radio quiet. Thus the origin of this jet emission is not conventional, i.e., may not be launched from the CENBOL. The physical reason behind 
these observations is discussed in \S 6. 

\subsubsection{X-ray}

While fitting a spectrum with the TCAF model, higher $N$ values are required as the source was active in radio i.e., had a jet. Using the constant normalization 
method of JCD17 and also using an additional PL model (for jet contribution in X-ray) with constant normalized TCAF model (for disk or inflow X-ray) 
as described in \S 3, we have estimated the X-ray flux contributions from jets/outflows ($F_{ouf}$) after separating accretion or inflowing disk flux ($F_{inf}$) 
contribution from total observed X-rays ($F_X$). The variation of the TCAF model fitted normalization ($N$) is shown in Fig. 1d and 2d. In Fig. 1e ad 2e, the variation 
of the $5.5$~GHz ATCA observed radio flux ($F_R$) is shown. The variation of $N$ is found to be roughly similar to the variation of $F_R$. This leads us to assume 
that higher $N$ is required to fit a spectrum, is due to additional X-ray flux contribution from the jets/outflows. During the entire outburst, we see a large 
variation of $N$ in the range of $0.51-3.10$. Interestingly, on the last observation on 2010 June 06 (MJD=55371.95), when the lowest $N$ value was obtained, 
$F_R$ was also observed in its very low values. This means that the entire X-ray flux ($F_X$) is contributed by the emission from the accretion disk or inflow alone 
and the X-ray contribution from jets/outflows towards the total observed X-ray may be neglected. Refitting the spectra with a frozen $N$ when it was lowest 
($=0.51$), allowed us to estimate X-ray flux contribution only from inflowing matter or accretion disk ($F_{inf}$). Using Eqn (2), we could estimate 
X-ray flux contribution from jets/outflows ($F_{ouf}$). We also estimated these X-ray fluxes using an additional PL model, considering jet X-ray 
follows power-law nature. The PL model was added with the TCAF model after freezing the TCAF model normalization at its minimum observed value (=$0.51$). 
The refit with the combined models, accounts the slight change in the TCAF model fitted parameters although within the acceptable limits (for more details see, Table 2 
of Paper I and Table 2 of the current paper). Here, `cflux' method calculated flux contributions from the TCAF and the PL models provide us $F_{inf}$ and $F_{ouf}$ 
respectively.

The variation of three different X-ray fluxes during 2009-10 outbursts of XTE~J1752-223 is shown in both Fig. 1(a-c) and Fig. 2(a-c) for only 
the TCAF and TCAF+PL models respectively. Using the first method, we see that $F_{ouf}$ reached to its maximum value on 2010 January 19 
(MJD = 55215.91) in HIMS. This was also the case for second method. There was no radio observation on that day. Maximum $F_R$ was observed during 
the immediate next observation of $5.5$~GHz ATCA data, $\sim 2$~days later (see, Fig 1e, 2e). From using only the TCAF model, $F_{ouf}$ is observed 
to decrease rapidly until MJD=55224.36 after which it remained almost constant for the next $\sim 16$~days (MJD=55240.01), before decreasing further into 
lower values during the declining phase of the outburst. We also calculated the percentage of jet X-ray flux ($F_{ouf}$) from the total X-ray ($F_X$) and 
see that the contribution of $F_{ouf}$ was maximum ($\sim 82~\%$) on the 2nd observation ID on 2009 November 2 (MJD = 55137.23). During SIMS or SS, the jet 
X-ray contribution to the total X-ray was high. However, using the second method, we see more variations of $F_{ouf}$, although the maximum flux 
occurred at the same date. Variation of $F_{ouf}$ is more analogous to the variation of radio flux ($F_R$) for the second method.

\subsection{Radio and X-ray Correlation}

Generally, radiation from accretion disk dominates in X-rays, while that from the jet dominates in radio. Since according to the TCAF solution, jets are 
launched from CENBOL, Radio and X-ray correlation indicates a coupling between disk and jet (Hannikainen et al. 1998; Corbel et al. 2000; 2003). The 
outflow rate also depends on the spectral nature i.e., the inflow rate of the two components of the accretion flows. So essentially, if the radio is high, 
then the jet X-ray and disk X-ray fluxes are also higher. Thus, clearly the halo rate is very high making it impossible to cool the CENBOL by soft photons 
from the Keplerian disk.  

In Fig. 3 and 4, we show the correlation plots between $F_R$ and X-ray fluxes ($F_X$, $F_{inf}$, $F_{ouf}$) for both our used methods respectively. 
The correlation plots are fitted using the relation $F_R \propto F_X^b$, where $b$ is the correlation index. The exact relation $y=f(x)=a~x^b$ is used 
in $gnuplot$ for the fitting. Here, an extra constant parameter `$a$' is used for equating left and right hand sides of the equation. 
For all correlations, we make use of the $5.5$~GHz radio data from Brocksopp et al. (2013). Using first method in Fig. 3(a), we show the correlation 
between radio ($F_R$) with the jet X-ray ($F_{ouf}$ in $2.5-25$~keV) and obtained a correlation index $b \sim 0.59 \pm 0.25$. In Fig. 3(b), the relation 
of $F_R$ with $F_{inf}$ (in $2.5-25$ keV) is shown, which follows a correlation with index $b \sim 0.76 \pm 0.23$. In Fig. 3(c), we show the relation of 
$F_R$ with $3-9$ keV X-ray flux ($F_X$), which follows a steeper correlation with $b \sim 1.28 \pm 0.28$. We also estimated the correlation between $F_R$ 
with total X-ray flux ($F_X$) in $2.5-25$ keV range. This is shown in Fig. 3(d) and we find that for this $b \sim 0.99 \pm 0.32$. 
Using second method in Fig. 4(a), we show the correlation between radio ($F_R$) with the jet X-ray ($F_{ouf}$ in $2.5-25$~keV) and obtained a 
correlation index $b \sim 0.71 \pm 0.23$. In Fig. 4(b), the correlation of $F_R$ with $F_{inf}$ (in $2.5-25$ keV) is shown, which follows a correlation with 
index $b \sim 0.43 \pm 0.23$. In Fig. 4(c) and 4(d), we show the correlation of $F_R$ with total X-ray fluxes in $3-9$ keV and $2.5-25$ keV bands respectively. 
Similar to earlier method, here we also observed the $F_R$ follows a steeper correlation with the X-ray fluxes in the above mentioned two energy bands 
i.e., $b \sim 1.08 \pm 0.28$, and $1.01 \pm 0.39$ respectively. In Table 3, we have listed all the correlation coefficients and indices for both the 
methods used.

From all the panels in Fig. 3 and 4, we see that although a good correlation exists between radio and X-ray fluxes, in high intensity regions of the outburst 
the points are scattered. More precisely, we see a tight correlation in the HS whereas a weak correlation (deviation of the observed/estimated points from 
the correlation curves) in the other states (HIMS, SIMS or SS). We may conclude that the jet is compact in the hard state and discrete in other states.

\section{Discussion and Concluding Remarks}

In Paper I, we described the accretion flow properties of the Galactic short orbital period transient BHC XTE~J1752-223 during its 2009-10 outburst based 
on spectral and timing analysis using RXTE/PCA, Swift/BAT and MAXI/GSC data. A detailed study was done using archival RXTE PCA (PCU2) data. Spectra are 
fitted with two types of models: i) the phenomenological DBB+PL model and ii) physical TCAF model. During the entire outburst phase (including softer states 
of SIMS or SS), the source was highly active in radio, i.e., in jets. This motivated us to find X-ray contribution from jets/outflows in the observed total 
X-rays using the method presented in JCD17. Similar to JCD17 who studied 2005 outburst of Swift~J1753.5-0127, we require higher normalization $N$ to fit 
spectra with the TCAF model in the high jet dominated regions. Besides the method of JCD17, we also used another method to estimate the X-ray flux 
contribution from the jet. We used an additional PL model to account for the contribution from the jet spectra when the TCAF normalization ($N$) was kept 
frozen to its minimum observed value. This minimum $N$ was observed when all model parameters of the TCAF model were kept, free while fitting spectra and 
results were presented in the Paper I.

There is a basic difference between the normalization of the TCAF model and other inbuilt models in XSPEC. In TCAF, the model normalization is a constant 
factor that is required to match the observed spectra with the theoretical one. In phenomenological models, it is customary to adjust the normalization for 
each observed data. However, since in TCAF, the shape of the entire spectrum comes at a time, the factor is supposed to remain constant across the spectral 
states which are observed with a particular satellite instrument. In the presence of jets/outflows, one could see a significant variation of $N$ as the 
current version of the TCAF model fits file, the X-rays emitted from the base of the jet are not included. During the entire 2009-10 outburst of XTE~J1752-223, 
a variation of $N$ in the range of $0.51-3.10$ was observed. When we compare its variation with that of $F_R$, we see a similar variation (see, Fig. 1d,e 
and Fig. 2d,e). Interestingly, we required the lowest $N$ value of $=0.51$  on the last observation day (2010 June 06 i.e., MJD=55371.95), when $F_R$ was 
also at its lowest value. One can assume that on this observation entire X-ray ($F_X$) was contributed by the emission from the inflowing matter alone (JCD17). 
This allowed us to estimate the X-ray flux contribution only from inflowing matter or accretion disk ($F_{inf}$) by refitting spectra with the frozen $N$ 
values at its lowest observed value ($=0.51$). Now, jet X-ray contribution i.e., $F_{ouf}$ was estimated in each observations from both our mentioned methods 
in \S3. Overall, we see a maximum of $82.68 \% $ with an average of $\sim 43.68 \% $ contribution of X-rays from jets to the total observed X-rays. 
This suggests that the source is jet dominated, especially in the intermediate spectral states. From the second method, we also estimated the total, inflow 
and outflow X-ray fluxes. Using this method, we observed a maximum of $86.78 \%$ with an average of $37.08 \%$ jet flux within the total observed X-ray flux. 
A small variations are also observed for the inflow and the outflow X-ray fluxes between the two methods. In Figure 1(b-c) and 2(b-c), although we see 
significant changes in inflow and outflow in the initial rising phase i.e., HS (Ris.), in other region of the outburst changes are insignificant. 
In the initial rising HS, we observed lower inflow rate and higher outflow rate with the first method, whereas with the second method, opposite features 
between the two fluxes are observed. In both these methods, the total X-ray flux showed almost similar variation throughout the entire outburst. 
The $F_{ouf}$ from both the methods shows roughly similar variations with the TCAF normalization ($N$) and radio flux ($F_R$). However, the variation 
of $F_{ouf}$ using the second method is more analogous to the variations of $N$ and $F_R$.

Comparative variations of the three types of X-ray fluxes with $N$ and $F_R$ are shown in Fig. 1 and Fig. 2. Maximum outflow flux ($F_{ouf}$) is observed 
on 2010 Jan 19 (MJD=55215.91), when the source was rediscovered after the Sun constraint period of the RXTE PCA was over. This maximum flux is observed on 
the same date from both the methods. Similarly, we saw a maximum of $F_R$ in the $5.5$~GHz ATCA data, when it was observed after the Sun constraint period. 
Due to non-observation of the radio on MJD=55215.91, we observed the maximum $F_R$ almost $2$~days later. Interestingly, on MJD=55215.91, we do not see 
maximum inflow flux $F_{inf}$, and it showed its peak flux on the HIMS to SIMS/SS transition day (2010 Jan. 22; MJD=55218.8). On this particular day, 
as $F_{inf}$ was higher, the cooling rate was higher to reduce the size of the CENBOL as well as its temperature (to make $R\sim 1$). This makes the spectral 
state softer. This is why we see a harder to softer spectral state transition on this particular day (MJD=55218.8). However, using additional power-law, 
we see maximum $F_{inf}$ on MJD 55220.7, which is one observation later than the case in the first method. After that, using only the TCAF model, we notice 
that all fluxes are reduced, before showing a marginal increasing trend during declining HIMS. During the declining HS, we see a reduction in all fluxes 
as well as $F_R$, since supply from the companion is probably blocked near the outer edge of the disk. However, the $F_{inf}$ and $F_{ouf}$ estimated with 
the second method, showed rise and dip natures in the declining phase (SIMS or SS state), what we also saw in the variation of $N$, and $F_R$.

Since we have been able to separate the total observed X-rays into its two components using TCAF model, we studied the correlation of the $F_R$ with the 
three types of X-ray fluxes ($F_X$, $F_{inf}$, $F_{ouf}$) in the form of $F_R \sim F_X^b$ (where $b$ is the correlation index), estimated with the two 
methods of \S3. Although, a steeper correlation is followed between $F_R$ and $F_X$ ($3-9$~keV) (Fig. 3c, 4c) using both methods, it is not the case for 
$F_{inf}$. While it was steeper using method 1, $F_R$ and $F_{inf}$ has been found to correlate weakly using the second method (Fig. 3b, 4b). 
$F_R$ and $F_{ouf}$ shows close correlation for both the methods. While the correlation was not so steeper in the first method, it has become steeper 
for the second method (Fig. 3a, 4a). $F_R$ and $F_X$ (in 2.5--25 keV) showed (Fig. 3d, 4d) steeper correlations for both the methods. This nature of 
the correlations are similar to other short orbital period transients BHCs Swift~J1753.5-0127 (JCD17), XTE~J1118+480 (Chatterjee et al. 2019; 
Debnath et al. 2020), MAXI~J1836-194 (Jana et al. 2020). These objects are defined as `outlier' as these sources do not show the standard correlation, 
when $F_R$ was correlated with $F_X$, measured in $3-9$~keV band. But, our analysis of these groups of short orbital period BHCs (including the present 
source) show `standard' correlation when $F_R$ is being correlated with X-ray flux of only from jets, i.e., with $F_{ouf}$. However, to firmly confirm 
this, we need more samples. We have listed the correlation coefficients and indices ($a$ and $b$) in Table 3.

To study the strength of the correlations between radio and the different component of X-ray fluxes from the statistical point of views, we make use 
of the Pearson Linear and Spearman Rank correlation methods. For $F_R$ vs $F_{ouf}$, we have found the coefficient values of $\sim 0.598$ and $\sim 0.672$ 
for the Pearson ($p$) and Spearman ($s$) methods respectively, where $F_{ouf}$ is estimated from spectral analysis using only the TCAF model. Roughly similar 
$p$ and $s$ values of $\sim 0.697$ and $0.705$ respectively are obtained when outflow flux is estimated from the spectral fit with the TCAF+PL models. 
This tells us that $F_R$ and $F_{ouf}$ are strongly correlated with each other and also supports the fact that $F_R$ vs $F_{ouf}$ 
correlation falls within the `standard' correlation range of $0.6-0.7$. In Table 3, we present the $p$ and $s$ coefficients for four sets of correlations 
between radio ($F_R$) and X-ray fluxes ($F_{ouf}$, $F_{inf}$, and $F_X$ in two bands), where X-ray flues are obtained from two types of methods in \S3. 
All the correlations show strong correlation coefficients except for the $p$ coefficient of $F_R$ vs $F_{inf}$ from the second method. 

Further, in all the four correlation plots of Fig. 3 and Fig. 4, we see tight correlations in the low intensity HS and weak correlations in the 
intermediate or softer spectral states (HIMS, SIMS or SS). Tight correlations imply that the nature of the jet is compact. Weaker correlations seen 
when X-ray and radio intensities were high could indicate the jet to be discrete or blobby. This result is consistent with the previous report 
(Jana et al. 2017). Theoretically, compact jets are thermal pressure-driven when the compression ratio $R$ is higher. Blobby jets are radiation 
pressure-driven as they are observed in the intermediate or softer states when shock becomes weaker (Chakrabarti 1999a,b). Also, the nature of 
the jets could be blobby when the optical depth of the base of the jet is high and the flow separates as blobs (Chakrabarti et al. 2001). 
In this case, the correlation between the radio and X-ray fluxes breaks down. This is what we see during the present outburst of XTE~J1752-223. 
As the outburst progresses, we see a rise in $F_{inf}$ and movement of the source towards intermediate or softer spectral states. Since the outflow 
rate and its nature is controlled by the compression ratio ($R$), we do not see a similar variation of $F_{ouf}$ and $F_{inf}$. Theoretically, 
the maximum outflow rate ($F_{ouf}$) could be seen in the intermediate shock strength, i.e., in the intermediate states (see, Chakrabarti 1999a). 
This is what we see during the present outburst as well as earlier studied two BHCs (Swift~J1753.5-0127 and MAXI~J1836-194) by our group 
(Jana et al. 2017, 2020). 

The outflow is generally absent in the SS. However, during the present outburst, we see significant outflows in the SS. This indicates that the physical 
processes responsible for this jet are different from what we see during hard and intermediate spectral states. This discrete jet is perhaps radiation 
pressure driven. Physically, the disk could be magnetically dominated in SS, when a large amount of matter (i.e., high accretion rate) is being accreted 
by the BH from its companion, which brings in a large amount of stochastic magnetic field. Due to azimuthal velocity, it forms toroidal flux tubes. 
There is very strong magnetic tension acting on these flux tubes. Due to very high magnetic field, magnetic tension becomes the dominant force which 
collapses the toroidal flux tubes. As a consequence of the collapse of the toroidal magnetic flux tubes, a large amount of matter may be removed as 
outflow in the transverse direction to the disk. This is known as the magnetic rubber band effect as suggested by Nandi et al. (2001). They suggested 
that this evacuation of matter towards the transverse direction of the disk is the reason for blobby components of jets/outflows causing soft X-ray dips. 
This also could manifest themselves as flares. We think that during the softer states (SIMS and SS) of the present outburst of XTE~J1752-223, the situation 
could be similar. The disk was magnetically dominated and the jet was launched from the outer disk. In the softer states, due to high accretion rates, 
Keplerian disk had to eject huge amount of matter along magnetic fields to remove most of the angular momentum (Blandford \& Payne 1982). So, the launching 
location of this jet may not be from the CENBOL as is possibly the case in a normal scenario. 

\section*{Acknowledgements}

This work made use of PCA data of NASA's RXTE satellite; radio data of ATCA observatory from Brocksopp et al. (2013) paper. 
K.C. acknowledges support from DST/INSPIRE (IF170233) fellowship.
D.D. and S.K.C. acknowledge partial supports from Department of Higher Education, Government of West Bengal, India and 
ISRO sponsored RESPOND project (ISRO/RES/2/418/17-18) fund.
D.C. and D.D. acknowledge support from DST/SERB sponsored Extra Mural Research project (EMR/2016/003918) fund.
A.J. and D.D. acknowledge support from DST/GITA sponsored India-Taiwan collaborative project (GITA/DST/TWN/P-76/2017) fund.

\section*{DATA AVAILABILITY}
We used archival data of {\it RXTE} PCA for this work. Radio data is used from Brocksopp et al. (2013).

\begin{figure*}
\vskip 1.5cm
  \centering
  \begin{minipage}[b]{0.45\textwidth}
    \includegraphics[width=\textwidth]{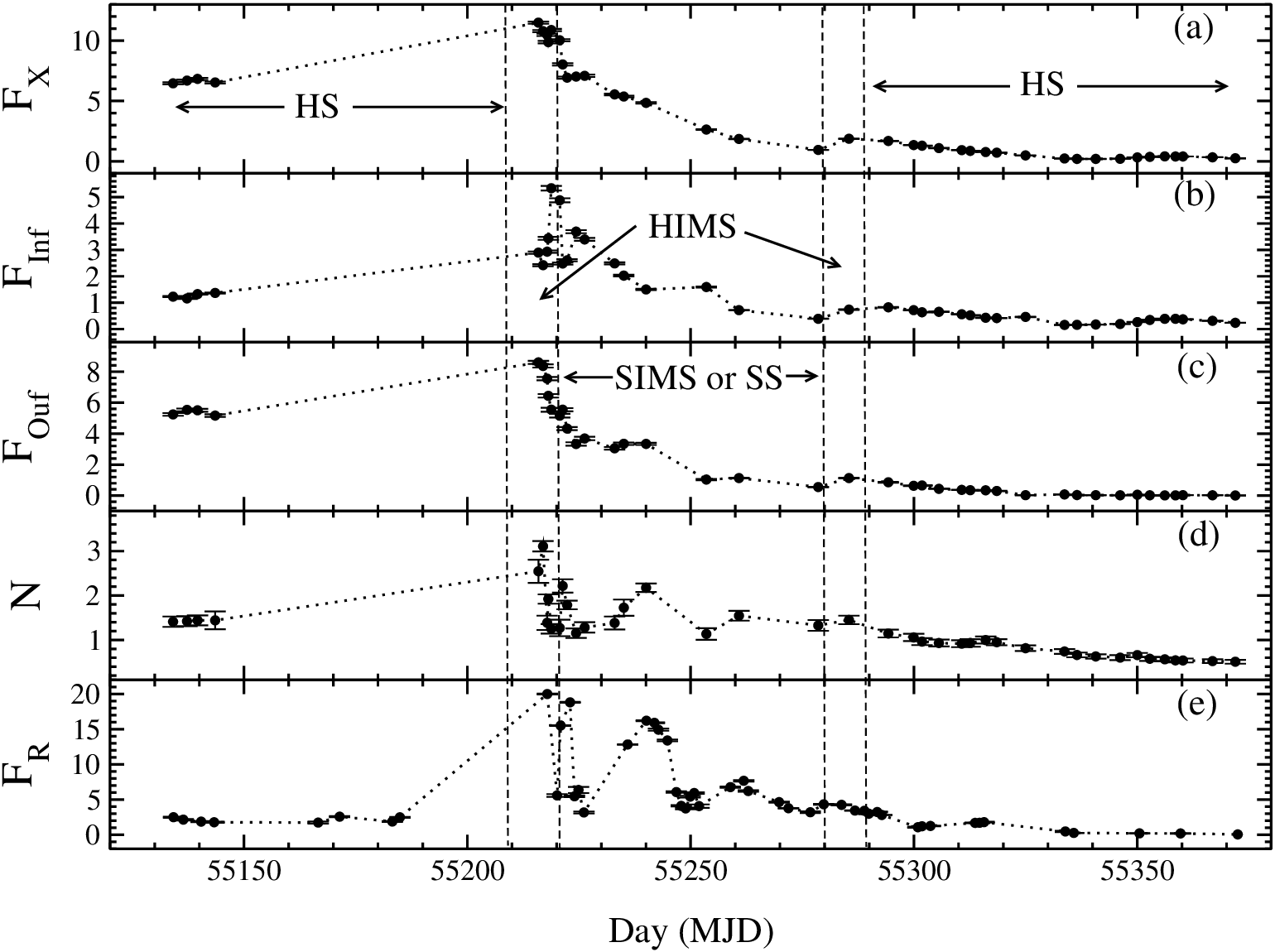}
    \caption{Variations of only TCAF model fitted (a) total X-ray flux ($F_X$), (b) accretion disk (inflow) X-ray flux ($F_{inf}$), (c) jet 
             (outflow) X-ray flux ($F_{ouf}$), (d) TCAF model fitted normalization ($N$) and (e) 5.5 GHz radio flux of ATCA (in $mJy$) with 
             time (day in MJD) are shown. All the X-ray fluxes ($F_X,~F_{inf},~F_{ouf}$) are shown in units of $10^{-9}~ erg~ cm^{-2}~ sec^{-1}$.}
\end{minipage}
  \hfill
  \begin{minipage}[b]{0.47\textwidth}
    \includegraphics[width=\textwidth]{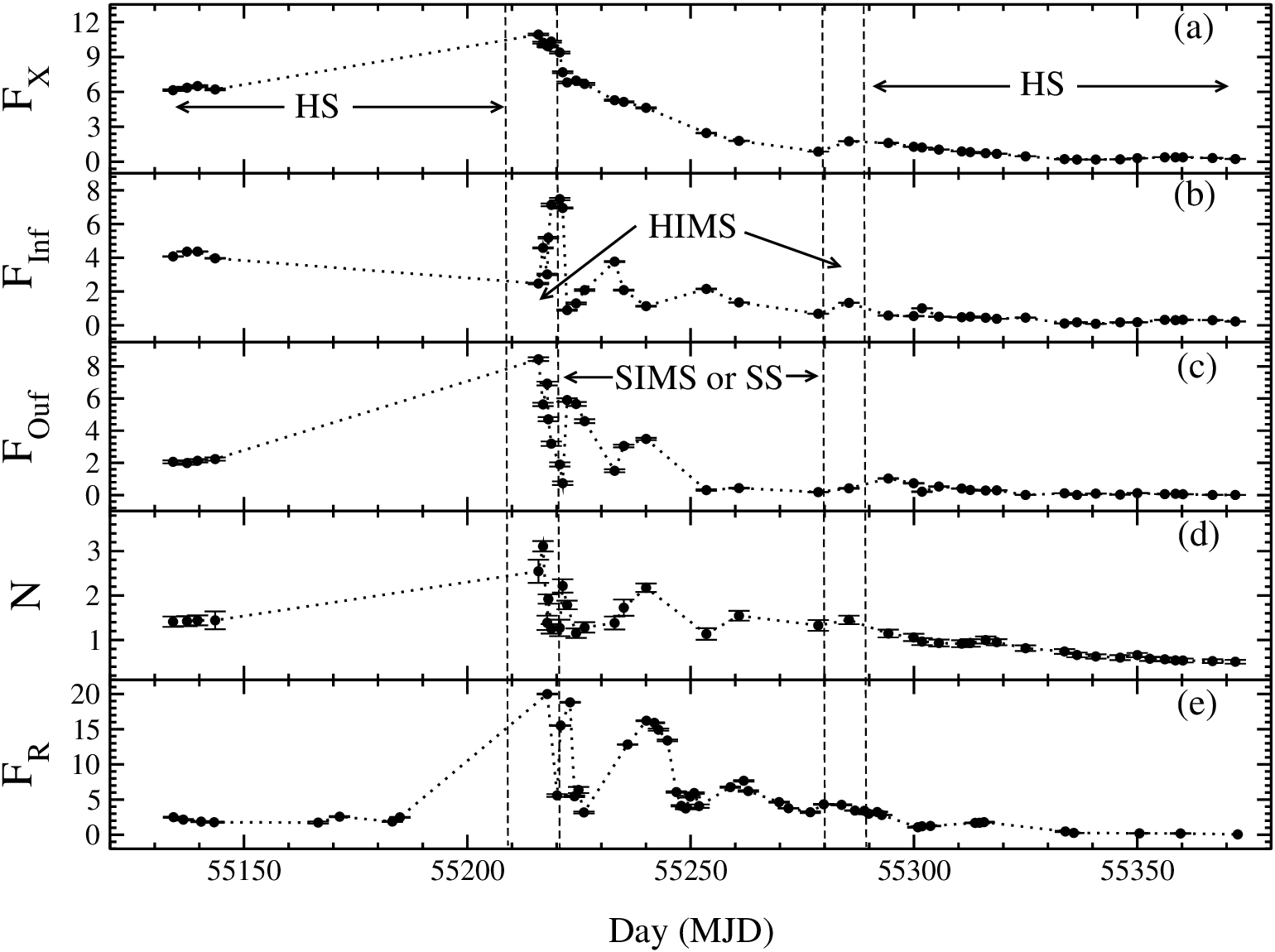}
    \caption{Variations of TCAF+power-law model fitted (a) total X-ray flux ($F_X$), (b) accretion disk (inflow) X-ray flux ($F_{inf}$), (c) 
             jet (outflow) X-ray flux ($F_{ouf}$), (d) TCAF model fitted normalization ($N$) and (e) 5.5 GHz radio flux of ATCA (in $mJy$) with 
             time (day in MJD) are shown. All the X-ray fluxes ($F_X,~F_{inf},~F_{ouf}$) are shown in units of $10^{-9}~ erg~ cm^{-2}~ sec^{-1}$.}
\end{minipage}
\end{figure*}

\begin{figure*}
\vskip 1.5cm
  \centering
  \begin{minipage}[b]{0.45\textwidth}
    \includegraphics[width=\textwidth]{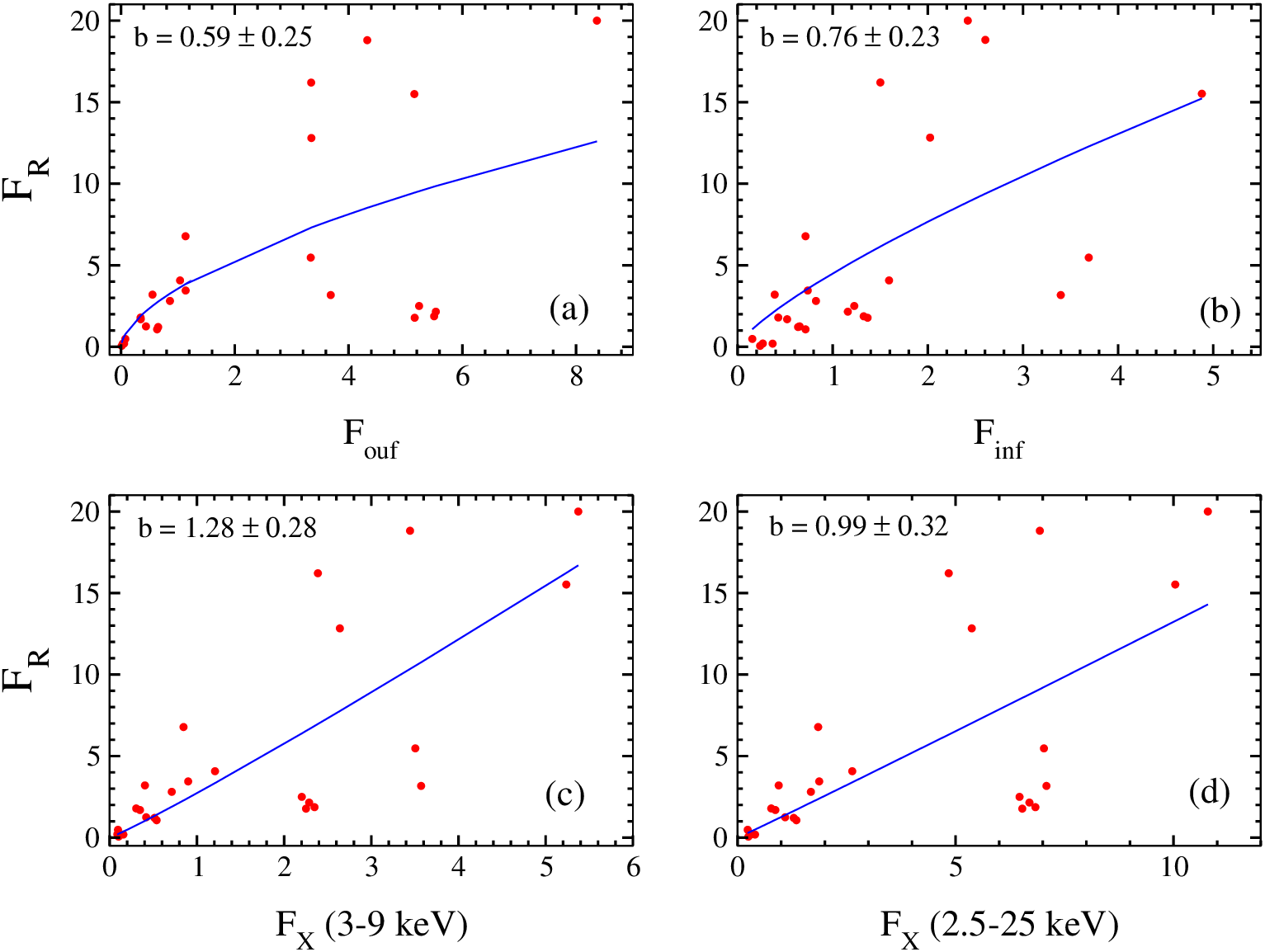}
    \caption{Correlation plots of radio ($F_R$) with (a) $2.5-25$ keV outflow X-ray ($F_{ouf}$), (b) $2.5-25$ keV inflow X-ray ($F_{inf}$),
          (c) $3-9$ keV total X-ray ($F_X$) and (d) $2.5-25$ keV total X-ray ($F_X$) fluxes. Radio data is taken from Brocksopp et al.
          (2013). All the X-ray fluxes are estimated by freezing TCAF normalization to the lowest value.}
\end{minipage}
  \hfill
  \begin{minipage}[b]{0.47\textwidth}
    \includegraphics[width=\textwidth]{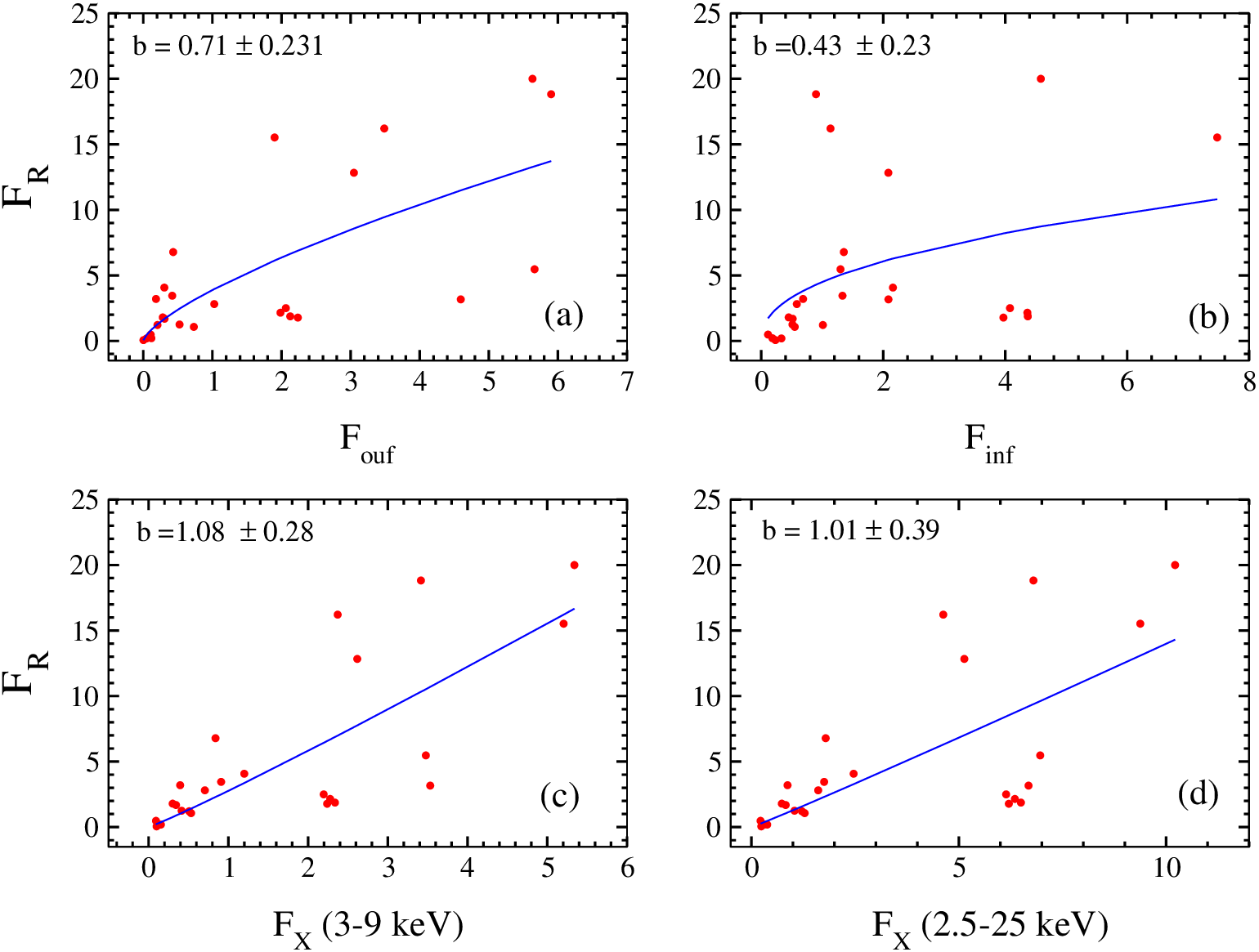}
    \caption{Correlation plots of radio ($F_R$) with (a) $2.5-25$ keV outflow X-ray ($F_{ouf}$), (b) $2.5-25$ keV inflow X-ray ($F_{inf}$),
          (c) $3-9$ keV total X-ray ($F_X$) and (d) $2.5-25$ keV total X-ray ($F_X$) fluxes. Radio data is taken from Brocksopp et al.
          (2013). All the X-ray fluxes are estimated using minimum TCAF normalization plus power-law models.}
\end{minipage}
\end{figure*}

\clearpage 
\begin{table*}
\small
 \centering
 \addtolength{\tabcolsep}{-2.5pt}
 \caption{Jet properties using only TCAF model}
 \label{tab:table1}
 \begin{tabular}{|c|c|c|c|c|c|c|c|c|}
 \hline

Obs ID & UT$^{[1]}$ &  MJD & N$^{[2]}$ & $F_X $$^{[3]}$ &  $F_{inf}$$^{[3]}$ &  $F_{ouf}$$^{[3]}$ & $F_X$$^{[3]}$ &  $F_{ouf}$$^{[4]}$ \\
 &  &  &  & (2.5-25 keV)  & (2.5-25 keV) & (2.5-25 keV) & (3-9 keV) & percent. \\
 (1) & (2) & (3) & (4) & (5) & (6) & (7) & (8) & (9) \\
\hline

94331-01-02-00  & 2009-10-30 & 55134.11  &  1.41 ${\pm~0.11}$ & 6.467  ${\pm~ 0.085}$ & 1.227 ${\pm ~0.019}$ & 5.240 ${\pm ~0.087}$ & 2.203 ${\pm~0.028 }$& 81.02\\
94331-01-02-06  & 2009-11-02 & 55137.23  &  1.42 ${\pm~0.11}$ & 6.692  ${\pm~ 0.088}$ & 1.158 ${\pm ~0.018}$ & 5.534 ${\pm ~0.090}$ & 2.287 ${\pm~0.029 }$& 82.68\\
94331-01-02-10  & 2009-11-04 & 55139.58  &  1.44 ${\pm~0.11}$ & 6.830  ${\pm~ 0.090}$ & 1.325 ${\pm ~0.020}$ & 5.505 ${\pm ~0.092}$ & 2.348 ${\pm~0.030 }$& 80.59\\
94331-01-03-05  & 2009-11-08 & 55143.53  &  1.44 ${\pm~0.19}$ & 6.530  ${\pm~ 0.086}$ & 1.367 ${\pm ~0.021}$ & 5.162 ${\pm ~0.089}$ & 2.251 ${\pm~0.028 }$& 79.05\\
94331-01-06-00  & 2010-01-19 & 55215.91  &  2.54 ${\pm~0.26}$ & 11.492 ${\pm~ 0.091}$ & 2.894 ${\pm ~0.045}$ & 8.597 ${\pm ~0.102}$ & 5.199 ${\pm~0.066 }$& 74.80\\
94331-01-06-01  & 2010-01-20 & 55216.95  &  3.10 ${\pm~0.11}$ & 10.787 ${\pm~ 0.092}$ & 2.419 ${\pm ~0.038}$ & 8.367 ${\pm ~0.100}$ & 5.374 ${\pm~0.068 }$& 77.56\\
94331-01-06-02  & 2010-01-21 & 55217.87  &  1.38 ${\pm~0.16}$ & 10.482 ${\pm~ 0.098}$ & 2.933 ${\pm ~0.046}$ & 7.548 ${\pm ~0.108}$ & 5.465 ${\pm~0.070 }$& 72.01\\
95360-01-01-08  & 2010-01-22 & 55218.14  &  1.92 ${\pm~0.10}$ & 9.877  ${\pm~ 0.100}$ & 3.435 ${\pm ~0.054}$ & 6.441 ${\pm ~0.114}$ & 5.325 ${\pm~0.068 }$& 65.21\\
95360-01-01-00  & 2010-01-22 & 55218.80  &  1.25 ${\pm~0.10}$ & 10.883 ${\pm~ 0.093}$ & 5.337 ${\pm ~0.084}$ & 5.545 ${\pm ~0.126}$ & 5.697 ${\pm~0.073 }$& 50.95\\
95360-01-01-02  & 2010-01-24 & 55220.68  &  1.27 ${\pm~0.18}$ & 10.038 ${\pm~ 0.092}$ & 4.880 ${\pm ~0.077}$ & 5.157 ${\pm ~0.120}$ & 5.235 ${\pm~0.067 }$& 51.38\\
95360-01-01-10  & 2010-01-25 & 55221.35  &  2.21 ${\pm~0.14}$ & 8.032  ${\pm~ 0.096}$ & 2.484 ${\pm ~0.039}$ & 5.548 ${\pm ~0.103}$ & 4.057 ${\pm~0.052 }$& 69.07\\
95360-01-01-12  & 2010-01-26 & 55222.33  &  1.78 ${\pm~0.09}$ & 6.932  ${\pm~ 0.091}$ & 2.604 ${\pm ~0.041}$ & 4.328 ${\pm ~0.100}$ & 3.444 ${\pm~0.044 }$& 62.43\\
95360-01-01-14  & 2010-01-28 & 55224.36  &  1.15 ${\pm~0.10}$ & 7.025  ${\pm~ 0.092}$ & 3.691 ${\pm ~0.058}$ & 3.334 ${\pm ~0.109}$ & 3.504 ${\pm~0.044 }$& 47.46\\
95360-01-02-02  & 2010-01-30 & 55226.25  &  1.28 ${\pm~0.11}$ & 7.084  ${\pm~ 0.093}$ & 3.397 ${\pm ~0.053}$ & 3.686 ${\pm ~0.108}$ & 3.570 ${\pm~0.045 }$& 52.03\\
95360-01-03-00  & 2010-02-05 & 55232.98  &  1.38 ${\pm~0.14}$ & 5.541  ${\pm~ 0.073}$ & 2.487 ${\pm ~0.039}$ & 3.054 ${\pm ~0.083}$ & 2.699 ${\pm~0.034 }$& 55.11\\
95360-01-03-01  & 2010-02-08 & 55235.03  &  1.72 ${\pm~0.18}$ & 5.370  ${\pm~ 0.071}$ & 2.023 ${\pm ~0.032}$ & 3.346 ${\pm ~0.077}$ & 2.639 ${\pm~0.033 }$& 62.31\\
95360-01-04-02  & 2010-02-13 & 55240.01  &  2.17 ${\pm~0.09}$ & 4.842  ${\pm~ 0.064}$ & 1.500 ${\pm ~0.023}$ & 3.342 ${\pm ~0.068}$ & 2.388 ${\pm~0.030 }$& 69.00\\
95360-01-06-00  & 2010-02-26 & 55253.51  &  1.13 ${\pm~0.12}$ & 2.629  ${\pm~ 0.034}$ & 1.592 ${\pm ~0.025}$ & 1.036 ${\pm ~0.042}$ & 1.207 ${\pm~0.015 }$& 39.42\\
95360-01-07-00  & 2010-03-05 & 55260.81  &  1.54 ${\pm~0.11}$ & 1.848  ${\pm~ 0.024}$ & 0.714 ${\pm ~0.011}$ & 1.133 ${\pm ~0.026}$ & 0.847 ${\pm~0.010 }$& 61.33\\
95360-01-09-04  & 2010-03-23 & 55278.58  &  1.32 ${\pm~0.12}$ & 0.943  ${\pm~ 0.012}$ & 0.389 ${\pm ~0.006}$ & 0.553 ${\pm ~0.013}$ & 0.404 ${\pm~0.005 }$& 58.64\\
95360-01-10-04  & 2010-03-30 & 55285.44  &  1.45 ${\pm~0.09}$ & 1.872  ${\pm~ 0.024}$ & 0.737 ${\pm ~0.011}$ & 1.135 ${\pm ~0.027}$ & 0.899 ${\pm~0.011 }$& 60.61\\
95360-01-11-05  & 2010-04-08 & 55294.26  &  1.14 ${\pm~0.08}$ & 1.683  ${\pm~ 0.022}$ & 0.823 ${\pm ~0.013}$ & 0.859 ${\pm ~0.025}$ & 0.711 ${\pm~0.009 }$& 51.04\\
95360-01-12-03  & 2010-04-13 & 55299.95  &  1.05 ${\pm~0.08}$ & 1.348  ${\pm~ 0.017}$ & 0.714 ${\pm ~0.011}$ & 0.633 ${\pm ~0.021}$ & 0.538 ${\pm~0.006 }$& 47.01\\
95360-01-12-04  & 2010-04-15 & 55301.80  &  0.96 ${\pm~0.07}$ & 1.289  ${\pm~ 0.017}$ & 0.636 ${\pm ~0.010}$ & 0.653 ${\pm ~0.019}$ & 0.514 ${\pm~0.006 }$& 50.67\\
95702-01-01-03  & 2010-04-19 & 55305.58  &  0.93 ${\pm~0.07}$ & 1.091  ${\pm~ 0.014}$ & 0.652 ${\pm ~0.010}$ & 0.438 ${\pm ~0.017}$ & 0.418 ${\pm~0.005 }$& 40.20\\
95702-01-02-01  & 2010-04-24 & 55310.70  &  0.91 ${\pm~0.07}$ & 0.928  ${\pm~ 0.012}$ & 0.560 ${\pm ~0.008}$ & 0.368 ${\pm ~0.015}$ & 0.367 ${\pm~0.004 }$& 39.64\\
95702-01-02-03  & 2010-04-26 & 55312.60  &  0.93 ${\pm~0.08}$ & 0.867  ${\pm~ 0.011}$ & 0.520 ${\pm ~0.008}$ & 0.346 ${\pm ~0.014}$ & 0.347 ${\pm~0.004 }$& 39.94\\
95702-01-03-00  & 2010-04-30 & 55316.05  &  0.99 ${\pm~0.07}$ & 0.771  ${\pm~ 0.010}$ & 0.428 ${\pm ~0.006}$ & 0.343 ${\pm ~0.012}$ & 0.303 ${\pm~0.003 }$& 44.48\\
95702-01-03-02  & 2010-05-02 & 55318.55  &  0.95 ${\pm~0.06}$ & 0.716  ${\pm~ 0.009}$ & 0.415 ${\pm ~0.006}$ & 0.300 ${\pm ~0.011}$ & 0.289 ${\pm~0.003 }$& 41.99\\
95702-01-04-01  & 2010-05-08 & 55325.00  &  0.81 ${\pm~0.06}$ & 0.483  ${\pm~ 0.006}$ & 0.458 ${\pm ~0.007}$ & 0.025 ${\pm ~0.009}$ & 0.190 ${\pm~0.002 }$& 5.273\\
95702-01-05-03  & 2010-05-17 & 55333.71  &  0.74 ${\pm~0.05}$ & 0.231  ${\pm~ 0.003}$ & 0.155 ${\pm ~0.002}$ & 0.075 ${\pm ~0.003}$ & 0.094 ${\pm~0.001 }$& 32.78\\
95702-01-05-06  & 2010-05-20 & 55336.51  &  0.66 ${\pm~0.05}$ & 0.195  ${\pm~ 0.002}$ & 0.161 ${\pm ~0.002}$ & 0.033 ${\pm ~0.003}$ & 0.081 ${\pm~0.001 }$& 17.15\\
95702-01-06-02  & 2010-05-24 & 55340.71  &  0.62 ${\pm~0.05}$ & 0.196  ${\pm~ 0.002}$ & 0.171 ${\pm ~0.002}$ & 0.025 ${\pm ~0.003}$ & 0.080 ${\pm~0.001 }$& 12.98\\
95702-01-07-01  & 2010-05-30 & 55346.17  &  0.60 ${\pm~0.05}$ & 0.210  ${\pm~ 0.002}$ & 0.191 ${\pm ~0.003}$ & 0.019 ${\pm ~0.004}$ & 0.086 ${\pm~0.001 }$& 9.032\\
95702-01-07-03  & 2010-06-03 & 55350.02  &  0.65 ${\pm~0.04}$ & 0.319  ${\pm~ 0.004}$ & 0.265 ${\pm ~0.004}$ & 0.053 ${\pm ~0.005}$ & 0.087 ${\pm~0.001 }$& 16.82\\
95702-01-08-02  & 2010-06-09 & 55356.17  &  0.56 ${\pm~0.04}$ & 0.406  ${\pm~ 0.005}$ & 0.388 ${\pm ~0.006}$ & 0.017 ${\pm ~0.008}$ & 0.055 ${\pm~0.000 }$& 4.303\\
95702-01-09-00  & 2010-06-11 & 55358.57  &  0.54 ${\pm~0.04}$ & 0.404  ${\pm~ 0.005}$ & 0.386 ${\pm ~0.005}$ & 0.017 ${\pm ~0.007}$ & 0.157 ${\pm~0.002 }$& 4.237\\
95702-01-09-01  & 2010-06-13 & 55360.23  &  0.53 ${\pm~0.04}$ & 0.400  ${\pm~ 0.003}$ & 0.368 ${\pm ~0.003}$ & 0.031 ${\pm ~0.004}$ & 0.156 ${\pm~0.001 }$& 7.997\\
95702-01-10-00  & 2010-06-19 & 55366.85  &  0.52 ${\pm~0.04}$ & 0.332  ${\pm~ 0.002}$ & 0.310 ${\pm ~0.002}$ & 0.022 ${\pm ~0.003}$ & 0.132 ${\pm~0.001 }$& 6.798\\
95702-01-10-02  & 2010-06-24 & 55371.95  &  0.51 ${\pm~0.04}$ & 0.248  ${\pm~ 0.001}$ & 0.237 ${\pm ~0.002}$ & 0.011 ${\pm ~0.002}$ & 0.102 ${\pm~0.001 }$& 4.496\\

\hline
 \end{tabular}
\noindent{
 \leftline{$^{[1]}$ UT dates are in yyyy-mm-dd format.} 
 \leftline{$^{[2]}$ TCAF model fitted normalization parameter ($N$) is shown in column 4.} 
 \leftline{$^{[3]}$ Calculated X-ray fluxes (in $10^{-9}~erg~cm^{-2}~sec^{-1}$) using TCAF model normalization are shown in column 5-8.}
 \leftline{$^{[4]}$ Percentage of X-ray flux contribution from the jet to the total X-ray flux is shown in column 9.} 
 \leftline {Note: average values of 90\% confidence $\pm$ error values obtained using `err' task in XSPEC.}
}
\end{table*}

\begin{table*}
\small
 \centering
 \addtolength{\tabcolsep}{-4.0pt}
 \caption{TCAF model parameters and jet properties using TCAF (with min `$N$') plus power-law models}
 \label{tab:table1}
 \begin{tabular}{|c|c|c|c|c|c|c|c|c|c|c|c|c|c|}
 \hline

Obs ID$^{[1]}$ & UT$^{[2]}$ &  MJD & ${\dot m}_d$$^{[3]}$ & ${\dot m}_h$$^{[3]}$ & $X_s$$^{[3]}$ & $R$$^{[3]}$ & $M_{BH}$$^{[3]}$ & ${\chi}^2_{red}$$^{[4]}$ & $F_X $$^{[5]}$ &  $F_{inf}$$^{[5]}$ &  $F_{ouf}$$^{[5]}$ & $F_X$$^{[5]}$ &  $F_{ouf}$$^{[5]}$ \\

 &  &  &  & &  &  &  &  &(2.5-25 keV)  & (2.5-25 keV) & (2.5-25 keV) & (3-9 keV) & percent. \\
 (1) & (2) & (3) & (4) & (5) & (6) & (7) & (8) & (9)  &  (10)  &  (11)  &  (12)  &  (13)  &  (14) \\
\hline

X-02-00  & A-10-30 & 55134.11  &  $0.0011^{\pm 0.0002 }$ & $3.52 ^{\pm 0.60 }$ & $233.0 ^{\pm 2.4 }$ & $2.40 ^{\pm 0.33 }$ & $10.9 ^{\pm 0.3 }$ & 0.824 & $6.141 ^{\pm 0.075 }$ & $4.078 ^{\pm 0.018 }$ & $2.062 ^{\pm 0.097 }$ &  $2.195 ^{\pm 0.030 }$ & 33.58  \\
X-02-06  & A-11-02 & 55137.23  &  $0.0015^{\pm 0.0002 }$ & $3.49 ^{\pm 0.62 }$ & $233.8 ^{\pm 2.8 }$ & $2.35 ^{\pm 0.25 }$ & $11.7 ^{\pm 0.3 }$ & 0.721 & $6.350 ^{\pm 0.077 }$ & $4.364 ^{\pm 0.017 }$ & $1.986 ^{\pm 0.100 }$ &  $2.277 ^{\pm 0.031 }$ & 31.27  \\
X-02-10  & A-11-04 & 55139.58  &  $0.0010^{\pm 0.0001 }$ & $3.15 ^{\pm 0.59 }$ & $215.7 ^{\pm 2.6 }$ & $2.38 ^{\pm 0.31 }$ & $11.5 ^{\pm 0.3 }$ & 0.807 & $6.498 ^{\pm 0.079 }$ & $4.370 ^{\pm 0.019 }$ & $2.127 ^{\pm 0.102 }$ &  $2.337 ^{\pm 0.032 }$ & 32.73  \\
X-03-05  & A-11-08 & 55143.53  &  $0.0011^{\pm 0.0001 }$ & $3.14 ^{\pm 0.54 }$ & $226.6 ^{\pm 2.6 }$ & $2.59 ^{\pm 0.26 }$ & $11.7 ^{\pm 0.3 }$ & 0.791 & $6.207 ^{\pm 0.075 }$ & $3.970 ^{\pm 0.020 }$ & $2.236 ^{\pm 0.098 }$ &  $2.240 ^{\pm 0.031 }$ & 36.03  \\
X-06-00  & B-01-19 & 55215.91  &  $ 1.14  ^{\pm 0.10 }$  & $5.30 ^{\pm 0.55 }$ & $36.7  ^{\pm 0.9 }$ & $3.90 ^{\pm 0.37 }$ & $9.7  ^{\pm 0.2 }$ & 1.139 & $10.92 ^{\pm 0.080 }$ & $2.474 ^{\pm 0.042 }$ & $8.446 ^{\pm 0.113 }$ &  $5.159 ^{\pm 0.072 }$ & 77.34  \\
X-06-01  & B-01-20 & 55216.95  &  $ 1.50  ^{\pm 0.21 }$  & $6.29 ^{\pm 0.59 }$ & $34.4  ^{\pm 0.9 }$ & $3.64 ^{\pm 0.30 }$ & $9.3  ^{\pm 0.2 }$ & 1.511 & $10.21 ^{\pm 0.081 }$ & $4.584 ^{\pm 0.035 }$ & $5.631 ^{\pm 0.110 }$ &  $5.340 ^{\pm 0.075 }$ & 55.12  \\
X-06-02  & B-01-21 & 55217.87  &  $ 1.61  ^{\pm 0.19 }$  & $5.77 ^{\pm 0.50 }$ & $36.2  ^{\pm 1.1 }$ & $3.96 ^{\pm 0.27 }$ & $9.3  ^{\pm 0.2 }$ & 1.314 & $9.945 ^{\pm 0.086 }$ & $3.016 ^{\pm 0.043 }$ & $6.929 ^{\pm 0.120 }$ &  $5.433 ^{\pm 0.076 }$ & 69.67  \\
Y-01-08  & B-01-22 & 55218.14  &  $ 1.31  ^{\pm 0.18 }$  & $5.54 ^{\pm 0.49 }$ & $33.7  ^{\pm 0.9 }$ & $3.64 ^{\pm 0.23 }$ & $9.3  ^{\pm 0.2 }$ & 1.727 & $9.899 ^{\pm 0.088 }$ & $5.187 ^{\pm 0.050 }$ & $4.712 ^{\pm 0.126 }$ &  $5.231 ^{\pm 0.074 }$ & 47.60  \\
Y-01-00  & B-01-22 & 55218.80  &  $ 5.28  ^{\pm 0.70 }$  & $1.96 ^{\pm 0.32 }$ & $34.4  ^{\pm 0.8 }$ & $1.22 ^{\pm 0.12 }$ & $8.9  ^{\pm 0.2 }$ & 1.902 & $10.32 ^{\pm 0.082 }$ & $7.131 ^{\pm 0.079 }$ & $3.190 ^{\pm 0.139 }$ &  $5.675 ^{\pm 0.079 }$ & 30.90  \\
Y-01-02  & B-01-24 & 55220.68  &  $ 6.54  ^{\pm 0.61 }$  & $2.11 ^{\pm 0.24 }$ & $34.4  ^{\pm 1.0 }$ & $1.25 ^{\pm 0.23 }$ & $8.9  ^{\pm 0.2 }$ & 1.564 & $9.376 ^{\pm 0.081 }$ & $7.475 ^{\pm 0.072 }$ & $1.900 ^{\pm 0.133 }$ &  $5.204 ^{\pm 0.073 }$ & 20.27  \\
Y-01-10  & B-01-25 & 55221.35  &  $ 5.00  ^{\pm 0.35 }$  & $1.46 ^{\pm 0.07 }$ & $34.4  ^{\pm 0.9 }$ & $1.09 ^{\pm 0.20 }$ & $8.9  ^{\pm 0.2 }$ & 0.860 & $7.683 ^{\pm 0.084 }$ & $6.947 ^{\pm 0.036 }$ & $0.735 ^{\pm 0.114 }$ &  $4.054 ^{\pm 0.056 }$ & 9.57   \\
Y-01-12  & B-01-26 & 55222.33  &  $ 4.27  ^{\pm 0.39 }$  & $1.20 ^{\pm 0.06 }$ & $56.4  ^{\pm 0.8 }$ & $1.06 ^{\pm 0.28 }$ & $9.5  ^{\pm 0.2 }$ & 0.994 & $6.800 ^{\pm 0.080 }$ & $0.898 ^{\pm 0.038 }$ & $5.901 ^{\pm 0.111 }$ &  $3.416 ^{\pm 0.048 }$ & 86.78  \\
Y-01-14  & B-01-28 & 55224.36  &  $ 4.29  ^{\pm 0.38 }$  & $0.97 ^{\pm 0.07 }$ & $43.9  ^{\pm 0.7 }$ & $1.07 ^{\pm 0.27 }$ & $9.5  ^{\pm 0.2 }$ & 1.085 & $6.962 ^{\pm 0.081 }$ & $1.299 ^{\pm 0.054 }$ & $5.662 ^{\pm 0.121 }$ &  $3.477 ^{\pm 0.048 }$ & 81.33  \\
Y-02-02  & B-01-30 & 55226.25  &  $ 4.56  ^{\pm 0.38 }$  & $0.81 ^{\pm 0.04 }$ & $34.5  ^{\pm 0.8 }$ & $1.09 ^{\pm 0.18 }$ & $9.5  ^{\pm 0.2 }$ & 1.121 & $6.680 ^{\pm 0.082 }$ & $2.086 ^{\pm 0.050 }$ & $4.593 ^{\pm 0.119 }$ &  $3.531 ^{\pm 0.049 }$ & 68.76  \\
Y-03-00  & B-02-05 & 55232.98  &  $ 4.71  ^{\pm 0.37 }$  & $0.68 ^{\pm 0.06 }$ & $34.4  ^{\pm 0.9 }$ & $1.09 ^{\pm 0.13 }$ & $9.5  ^{\pm 0.2 }$ & 1.423 & $5.283 ^{\pm 0.064 }$ & $3.773 ^{\pm 0.036 }$ & $1.509 ^{\pm 0.092 }$ &  $2.684 ^{\pm 0.037 }$ & 28.57  \\
Y-03-01  & B-02-08 & 55235.03  &  $ 4.26  ^{\pm 0.40 }$  & $0.65 ^{\pm 0.04 }$ & $34.5  ^{\pm 0.8 }$ & $1.08 ^{\pm 0.21 }$ & $9.5  ^{\pm 0.3 }$ & 0.941 & $5.132 ^{\pm 0.062 }$ & $2.084 ^{\pm 0.029 }$ & $3.047 ^{\pm 0.086 }$ &  $2.616 ^{\pm 0.036 }$ & 59.38  \\
Y-04-02  & B-02-13 & 55240.01  &  $ 4.07  ^{\pm 0.18 }$  & $0.58 ^{\pm 0.04 }$ & $34.8  ^{\pm 0.9 }$ & $1.10 ^{\pm 0.17 }$ & $9.8  ^{\pm 0.2 }$ & 1.062 & $4.624 ^{\pm 0.056 }$ & $1.135 ^{\pm 0.022 }$ & $3.488 ^{\pm 0.075 }$ &  $2.370 ^{\pm 0.033 }$ & 75.44  \\
Y-06-00  & B-02-26 & 55253.51  &  $ 4.37  ^{\pm 0.19 }$  & $0.47 ^{\pm 0.03 }$ & $34.4  ^{\pm 0.9 }$ & $1.21 ^{\pm 0.12 }$ & $9.5  ^{\pm 0.2 }$ & 0.959 & $2.463 ^{\pm 0.030 }$ & $2.158 ^{\pm 0.023 }$ & $0.304 ^{\pm 0.047 }$ &  $1.200 ^{\pm 0.016 }$ & 12.36  \\
Y-07-00  & B-03-05 & 55260.81  &  $ 3.38  ^{\pm 0.24 }$  & $0.46 ^{\pm 0.05 }$ & $34.4  ^{\pm 0.9 }$ & $1.21 ^{\pm 0.17 }$ & $9.5  ^{\pm 0.2 }$ & 0.946 & $1.789 ^{\pm 0.021 }$ & $1.355 ^{\pm 0.010 }$ & $0.433 ^{\pm 0.029 }$ &  $0.840 ^{\pm 0.011 }$ & 24.24  \\
Y-09-04  & B-03-23 & 55278.58  &  $ 3.07  ^{\pm 0.21 }$  & $0.37 ^{\pm 0.04 }$ & $34.5  ^{\pm 0.9 }$ & $1.22 ^{\pm 0.12 }$ & $9.5  ^{\pm 0.3 }$ & 0.912 & $0.869 ^{\pm 0.010 }$ & $0.685 ^{\pm 0.005 }$ & $0.183 ^{\pm 0.015 }$ &  $0.396 ^{\pm 0.005 }$ & 21.11  \\
Y-10-04  & B-03-30 & 55285.44  &  $ 2.25  ^{\pm 0.20 }$  & $0.60 ^{\pm 0.28 }$ & $34.7  ^{\pm 1.7 }$ & $1.05 ^{\pm 0.17 }$ & $10.6 ^{\pm 0.3 }$ & 0.752 & $1.752 ^{\pm 0.021 }$ & $1.332 ^{\pm 0.010 }$ & $0.42  ^{\pm 0.030 }$ &  $0.909 ^{\pm 0.012 }$ & 23.96  \\
Y-11-05  & B-04-08 & 55294.26  &  $0.0017^{\pm 0.0002 }$ & $3.31 ^{\pm 0.34 }$ & $129.3 ^{\pm 1.6 }$ & $2.44 ^{\pm 0.35 }$ & $10.9 ^{\pm 0.3 }$ & 0.757 & $1.609 ^{\pm 0.019 }$ & $0.582 ^{\pm 0.012 }$ & $1.026 ^{\pm 0.028 }$ &  $0.705 ^{\pm 0.009 }$ & 63.79  \\
Y-12-03  & B-04-13 & 55299.95  &  $0.0017^{\pm 0.0002 }$ & $3.13 ^{\pm 0.37 }$ & $132.5 ^{\pm 1.8 }$ & $2.55 ^{\pm 0.25 }$ & $11.0 ^{\pm 0.2 }$ & 1.011 & $1.282 ^{\pm 0.015 }$ & $0.549 ^{\pm 0.010 }$ & $0.732 ^{\pm 0.023 }$ &  $0.533 ^{\pm 0.007 }$ & 57.11  \\
Y-12-04  & B-04-15 & 55301.80  &  $0.0018^{\pm 0.0002 }$ & $2.92 ^{\pm 0.39 }$ & $114.0 ^{\pm 1.9 }$ & $2.20 ^{\pm 0.24 }$ & $9.35 ^{\pm 0.2 }$ & 0.851 & $1.216 ^{\pm 0.014 }$ & $1.010 ^{\pm 0.009 }$ & $0.206 ^{\pm 0.021 }$ &  $0.509 ^{\pm 0.007 }$ & 16.93  \\
Z-01-03  & B-04-19 & 55305.58  &  $0.0017^{\pm 0.0002 }$ & $2.94 ^{\pm 0.38 }$ & $123.2 ^{\pm 1.9 }$ & $2.60 ^{\pm 0.26 }$ & $11.6 ^{\pm 0.3 }$ & 0.998 & $1.037 ^{\pm 0.012 }$ & $0.511 ^{\pm 0.009 }$ & $0.526 ^{\pm 0.019 }$ &  $0.415 ^{\pm 0.005 }$ & 50.68  \\
Z-02-01  & B-04-24 & 55310.70  &  $0.0017^{\pm 0.0001 }$ & $2.89 ^{\pm 0.30 }$ & $126.1 ^{\pm 1.9 }$ & $2.50 ^{\pm 0.28 }$ & $11.2 ^{\pm 0.3 }$ & 0.881 & $0.883 ^{\pm 0.010 }$ & $0.481 ^{\pm 0.008 }$ & $0.401 ^{\pm 0.016 }$ &  $0.364 ^{\pm 0.005 }$ & 45.45  \\
Z-02-03  & B-04-26 & 55312.60  &  $0.0017^{\pm 0.0001 }$ & $2.91 ^{\pm 0.35 }$ & $127.8 ^{\pm 1.4 }$ & $2.33 ^{\pm 0.26 }$ & $9.85 ^{\pm 0.3 }$ & 1.073 & $0.824 ^{\pm 0.010 }$ & $0.514 ^{\pm 0.007 }$ & $0.309 ^{\pm 0.015 }$ &  $0.344 ^{\pm 0.004 }$ & 37.53  \\
Z-03-00  & B-04-30 & 55316.05  &  $0.0018^{\pm 0.0001 }$ & $2.90 ^{\pm 0.31 }$ & $124.8 ^{\pm 2.2 }$ & $2.50 ^{\pm 0.28 }$ & $11.2 ^{\pm 0.3 }$ & 1.138 & $0.733 ^{\pm 0.008 }$ & $0.450 ^{\pm 0.006 }$ & $0.282 ^{\pm 0.013 }$ &  $0.301 ^{\pm 0.004 }$ & 38.53  \\
Z-03-02  & B-05-02 & 55318.55  &  $0.0018^{\pm 0.0001 }$ & $2.94 ^{\pm 0.31 }$ & $125.1 ^{\pm 1.5 }$ & $2.58 ^{\pm 0.21 }$ & $10.1 ^{\pm 0.3 }$ & 0.967 & $0.680 ^{\pm 0.008 }$ & $0.385 ^{\pm 0.006 }$ & $0.294 ^{\pm 0.012 }$ &  $0.286 ^{\pm 0.004 }$ & 43.29  \\
Z-04-01  & B-05-08 & 55325.00  &  $0.0018^{\pm 0.0001 }$ & $2.53 ^{\pm 0.36 }$ & $133.3 ^{\pm 1.8 }$ & $2.46 ^{\pm 0.24 }$ & $11.0 ^{\pm 0.2 }$ & 0.789 & $0.460 ^{\pm 0.005 }$ & $0.456 ^{\pm 0.006 }$ & $0.004 ^{\pm 0.010 }$ &  $0.189 ^{\pm 0.002 }$ & 0.93   \\
Z-05-03  & B-05-17 & 55333.71  &  $0.0018^{\pm 0.0002 }$ & $2.83 ^{\pm 0.39 }$ & $139.4 ^{\pm 1.9 }$ & $2.63 ^{\pm 0.33 }$ & $9.63 ^{\pm 0.3 }$ & 0.637 & $0.219 ^{\pm 0.002 }$ & $0.111 ^{\pm 0.002 }$ & $0.108 ^{\pm 0.004 }$ &  $0.094 ^{\pm 0.001 }$ & 49.18  \\
Z-05-06  & B-05-20 & 55336.51  &  $0.0018^{\pm 0.0002 }$ & $2.93 ^{\pm 0.30 }$ & $128.9 ^{\pm 1.7 }$ & $2.68 ^{\pm 0.32 }$ & $9.59 ^{\pm 0.3 }$ & 0.901 & $0.188 ^{\pm 0.002 }$ & $0.183 ^{\pm 0.002 }$ & $0.004 ^{\pm 0.004 }$ &  $0.080 ^{\pm 0.001 }$ & 2.44   \\
Z-06-02  & B-05-24 & 55340.71  &  $0.0018^{\pm 0.0002 }$ & $2.83 ^{\pm 0.26 }$ & $140.0 ^{\pm 1.8 }$ & $2.46 ^{\pm 0.18 }$ & $9.40 ^{\pm 0.3 }$ & 0.839 & $0.185 ^{\pm 0.002 }$ & $0.094 ^{\pm 0.002 }$ & $0.091 ^{\pm 0.004 }$ &  $0.079 ^{\pm 0.001 }$ & 49.13  \\
Z-07-01  & B-05-30 & 55346.17  &  $0.0018^{\pm 0.0002 }$ & $2.93 ^{\pm 0.33 }$ & $131.6 ^{\pm 1.7 }$ & $2.60 ^{\pm 0.35 }$ & $10.0 ^{\pm 0.3 }$ & 0.932 & $0.202 ^{\pm 0.002 }$ & $0.178 ^{\pm 0.002 }$ & $0.024 ^{\pm 0.004 }$ &  $0.086 ^{\pm 0.001 }$ & 12.17  \\
Z-07-03  & B-06-03 & 55350.02  &  $0.0018^{\pm 0.0001 }$ & $2.79 ^{\pm 0.36 }$ & $138.8 ^{\pm 1.9 }$ & $2.54 ^{\pm 0.25 }$ & $9.25 ^{\pm 0.2 }$ & 0.788 & $0.303 ^{\pm 0.003 }$ & $0.184 ^{\pm 0.003 }$ & $0.118 ^{\pm 0.006 }$ &  $0.127 ^{\pm 0.001 }$ & 39.16  \\
Z-08-02  & B-06-09 & 55356.17  &  $0.0018^{\pm 0.0001 }$ & $2.73 ^{\pm 0.36 }$ & $129.0 ^{\pm 2.0 }$ & $2.69 ^{\pm 0.26 }$ & $9.92 ^{\pm 0.3 }$ & 0.875 & $0.380 ^{\pm 0.004 }$ & $0.322 ^{\pm 0.005 }$ & $0.058 ^{\pm 0.008 }$ &  $0.154 ^{\pm 0.000 }$ & 15.33  \\
Z-09-00  & B-06-11 & 55358.57  &  $0.0018^{\pm 0.0001 }$ & $2.72 ^{\pm 0.28 }$ & $132.1 ^{\pm 1.9 }$ & $2.63 ^{\pm 0.28 }$ & $9.99 ^{\pm 0.2 }$ & 0.611 & $0.380 ^{\pm 0.004 }$ & $0.307 ^{\pm 0.005 }$ & $0.073 ^{\pm 0.009 }$ &  $0.156 ^{\pm 0.000 }$ & 19.34  \\
Z-09-01  & B-06-13 & 55360.23  &  $0.0017^{\pm 0.0001 }$ & $2.76 ^{\pm 0.33 }$ & $133.8 ^{\pm 1.4 }$ & $2.54 ^{\pm 0.26 }$ & $9.42 ^{\pm 0.3 }$ & 0.998 & $0.379 ^{\pm 0.004 }$ & $0.330 ^{\pm 0.005 }$ & $0.048 ^{\pm 0.008 }$ &  $0.155 ^{\pm 0.002 }$ & 12.85  \\
Z-10-00  & B-06-19 & 55366.85  &  $0.0018^{\pm 0.0002 }$ & $2.76 ^{\pm 0.30 }$ & $126.0 ^{\pm 2.3 }$ & $2.72 ^{\pm 0.28 }$ & $9.39 ^{\pm 0.3 }$ & 0.852 & $0.315 ^{\pm 0.002 }$ & $0.308 ^{\pm 0.003 }$ & $0.007 ^{\pm 0.005 }$ &  $0.130 ^{\pm 0.001 }$ & 2.22   \\
Z-10-02  & B-06-24 & 55371.95  &  $0.0018^{\pm 0.0002 }$ & $2.70 ^{\pm 0.22 }$ & $137.0 ^{\pm 1.9 }$ & $2.45 ^{\pm 0.19 }$ & $9.89 ^{\pm 0.2 }$ & 0.924 & $0.235 ^{\pm 0.002 }$ & $0.232 ^{\pm 0.002 }$ & $0.002 ^{\pm 0.004 }$ &  $0.101 ^{\pm 0.001 }$ & 1.19   \\

\hline
 \end{tabular}
\noindent{
 \leftline{$^{[2]}$ Column 1 represents the Obs. Ids used for this work, where `X', `Y' and `Z' stand for 94331-01, 95360-01 
                           and 95702-01 respectively.}
 \leftline{$^{[2]}$ UT dates are in mm/dd format. First 4 observations are from 2009 (A) and rest from 2010 (B).}
 \leftline{$^{[3]}$ Combined TCAF (with minimum normalization) + power-law model fitted parameters is shown in column 4-8.}
 \leftline{$^{[4]}$  TCAF + power-law model fitted $\chi^2_{red}$ is shown in column 9.}
 \leftline{$^{[5]}$ Calculated X-ray fluxes (in $10^{-9}~erg~cm^{-2}~sec^{-1}$) using TCAF + power-law model 10-13.}
 \leftline{Note: average values of 90\% confidence $\pm$ error values obtained using `err' task in XSPEC. The errors are written
                  as superscripts to save space.}
}
\end{table*}

\begin{table*}
 \centering
 \addtolength{\tabcolsep}{-2.5pt}
 \caption{Statistical Coefficients for Radio and X-ray correlations}
 \label{tab:table3}
 \begin{tabular}{|cccccc|}
 \hline
                  &                &      $F_{ouf}$    &      $F_{inf}$    &       $F_X$       &      $F_X$         \\
  Methods         &  Coefficients  &    (2.5--25 keV)  &   (2.5--25 keV)   &    (3--9 keV)     &  (2.5--25 keV)     \\
    (1)           &     (2)        &       (3)         &       (4)         &        (5)        &      (6)           \\
\hline           
                  &      a         & $ 3.57 \pm 1.45 $ & $ 4.52 \pm 1.17 $ & $ 2.71 \pm 1.06 $ & $ 1.34 \pm 1.02 $  \\
  Only TCAF       &      b         & $ 0.59 \pm 0.25 $ & $ 0.76 \pm 0.23 $ & $ 1.28 \pm 0.28 $ & $ 0.99 \pm 0.32 $  \\
 with min $N$     &      $p$       &    $  0.598  $    &    $  0.625  $    &    $  0.759  $    &   $  0.663  $      \\
                  &      $s$       &    $  0.672  $    &    $  0.795  $    &    $  0.809  $    &   $  0.744  $      \\            
\hline
                  &      a         & $ 3.85 \pm 1.29 $ & $ 4.49 \pm 1.35 $ & $ 2.72 \pm 1.06 $ & $ 1.35 \pm 1.03 $  \\
 TCAF with min    &      b         & $ 0.71 \pm 0.23 $ & $ 0.43 \pm 0.23 $ & $ 1.08 \pm 0.28 $ & $ 1.01 \pm 0.39 $  \\
 $N$ + power-law  &      $p$       &    $  0.697  $    &    $  0.394  $    &    $  0.759  $    &   $  0.665  $      \\
                  &      $s$       &    $  0.705  $    &    $  0.642  $    &    $  0.810  $    &   $  0.756  $      \\
\hline
 \end{tabular}
\noindent{
\leftline{`$a$' and `$b$' are the correlation coefficient and index for correlation between radio and X-ray fluxes respectively, where $F_R = a F_X^b$ relation is followed.} 
\leftline{$F_X$ is replaced by $F_{ouf}$, $F_{inf}$, $F_X$ (in 3--9 keV) and $F_X$ (in 2.5--25 keV) mentioned in Cols. 3, 4, 5, and 6 respectively for different set of correlations.} 
\leftline{`$p$' and `$s$' represent the Pearson Linear and Spearman Rank coefficients for the same set of correlations respectively.}}

\end{table*}

\end{document}